\documentclass[12pt]{article}
\usepackage{epsfig}

\begin{document}

\begin{center}

\vspace*{1.0cm}

{\Large \bf{First search for double $\beta$ decay of platinum by ultra-low background HP Ge $\gamma$ spectrometry}}

\vskip 1.0cm

{\bf
P.~Belli$^a$,
R.~Bernabei$^{a,b,}$\footnote{Corresponding author.
   {\it E-mail address:} rita.bernabei@roma2.infn.it (R.~Bernabei).}
F.~Cappella$^{c,d}$,
R.~Cerulli$^{e}$,
F.A.~Danevich$^{f}$,
A.~Di~Marco$^a$,
A.~Incicchitti$^{c,d}$,
M.~Laubenstein$^{e}$,
S.S.~Nagorny$^{f}$,
S.~Nisi$^{e}$,
O.G.~Polischuk$^{f}$,
V.I.~Tretyak$^{f}$
}

\vskip 0.3cm

$^a${\it INFN sezione Roma ``Tor Vergata'', I-00133 Rome, Italy}

$^b${\it Dipartimento di Fisica, Universit\`a di Roma ``Tor Vergata'', I-00133, Rome, Italy}

$^c${\it INFN sezione Roma, I-00185 Rome, Italy}

$^d${\it Dipartimento di Fisica, Universit\`a di Roma ``La Sapienza'', I-00185, Rome, Italy}

$^e${\it INFN, Laboratori Nazionali del Gran Sasso, I-67010 Assergi (Aq), Italy}

$^f${\it Institute for Nuclear Research, MSP 03680 Kyiv, Ukraine}

\end{center}

\vskip 0.5cm

\begin{abstract}
A search for double $\beta$ processes in $^{190}$Pt and
$^{198}$Pt was realized with the help of ultra-low background HP
Ge 468 cm$^3$ $\gamma$ spectrometer in the underground Gran Sasso
National Laboratories of the INFN (Italy). After 1815 h of data
taking with 42.5 g platinum sample, $T_{1/2}$ limits on $2\beta$ processes
in $^{190}$Pt ($\varepsilon\beta^+$ and $2\varepsilon$) have been
established on the level of $10^{14}-10^{16}$ yr, 3 to 4 orders of
magnitude higher than those known previously. In particular, a
possible resonant double electron capture in $^{190}$Pt was
restricted on the level of $2.9\times10^{16}$ yr at 90\% C.L. In addition,
$T_{1/2}$ limit on $2\beta^-$ decay of $^{198}$Pt ($2\nu+0\nu$) to
the 2$^+_1$ excited level of $^{198}$Hg has been set at the first
time: $T_{1/2}>3.5\times10^{18}$ yr. The radiopurity level of the
used platinum sample is reported.
\end{abstract}

\vskip 0.4cm


\vskip 0.4cm


\section{Introduction}

The investigation of the double beta ($2\beta$) decay is considered
now as one of the most sensitive probe of physics beyond the
Standard Model of particles and interactions \cite{DBD}.
The observation of the neutrinoless ($0\nu$) mode of $2\beta$ decay could
establish an absolute scale of the neutrino mass and the neutrino mass
hierarchy, clarify the nature of the neutrino (Dirac or Majorana
particle) and check the conservation of the lepton number. The
process also depends on possible existence of
light Nambu-Goldstone bosons (Majorons) and
hypothetical admixture of right-handed currents in weak interaction.

At present the two neutrino (2$\nu$) 2$\beta^-$ decay, process of
nuclear transformation $(A,Z) \to (A,Z+2) + 2e^- + 2\widetilde{\nu}_e$,
was observed
in 10 isotopes with the half-lives in the range of
$10^{18}-10^{24}$ yr, while only limits on $0\nu2\beta^-$ decays
up to $10^{23}-10^{25}$ yr were set in numerous experiments
\cite{Bar10,DBD-tab}.

The sensitivity of experiments to search for $2\beta$ processes, which
transform $(A,Z)$ nucleus to $(A,Z-2)$: double electron capture
($2\varepsilon$), electron capture with emission of posi\-t\-ron
($\varepsilon\beta^{+}$), and double positron ($2\beta^{+}$)
decay, are on the level of $10^{15}-10^{21}$ yr
\cite{DBD-tab,Gav06,Bar07a,Kim07,Bar08,Daw08,Bel08a,Bel08b,Bar09,Bel09a,Bel09b,Bel09c,Ruk10,Bel11};
even allowed in the Standard Model the two neutrino mode is still
not observed\footnote{An indication on the double $\beta$ decay of
$^{130}$Ba was obtained in \cite{Mes01} by geochemical method.
However, this result has to be confirmed by direct counting
experiments.}. At the same time, the investigation of neutrinoless
$2\varepsilon$ and $\varepsilon\beta^{+}$ processes could give an
important information about possible contribution of the
right-handed currents to the neutrinoless double $\beta^{-}$ decay
rate \cite{Hir94}. Therefore developments of experimental methods
to search for "double beta plus" processes are required.

The $^{190}$Pt is one of the twenty two potentially $\varepsilon\beta^+$
active nuclei \cite{DBD-tab}. The energy of the double beta decay is
$Q_{2\beta} =(1383\pm6)$ keV \cite{Aud03}. The decay scheme of
the $^{190}$Pt is presented in Fig.~1.

\begin{figure}[htb]
\begin{center}
\resizebox{0.55\textwidth}{!}{\includegraphics{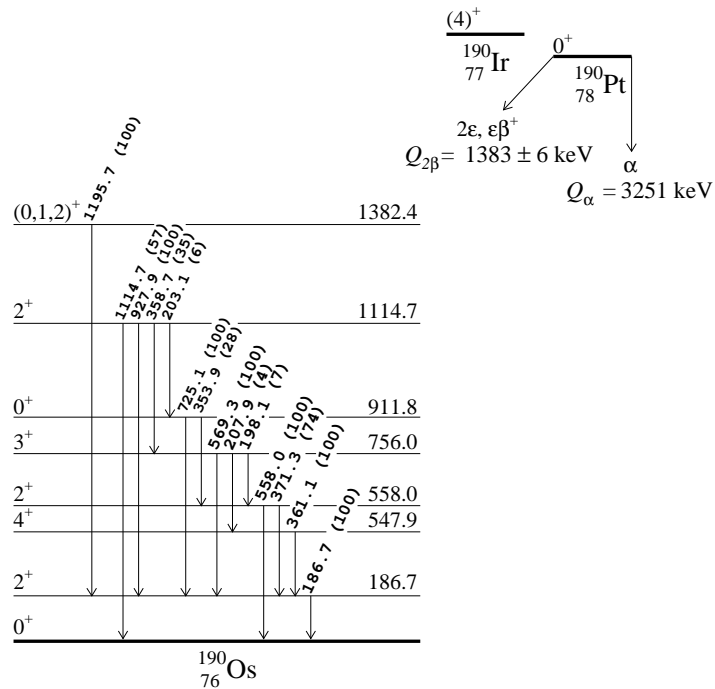}}
\end{center}
\caption{The decay scheme of the $^{190}$Pt \cite{Sin03}. The energies of the
excited levels and of the emitted $\gamma$ quanta are in keV (the relative
intensities of the $\gamma$ quanta are given in parentheses).
$Q_{2\beta}$ is the double beta decay energy.}
\end{figure}

It is also worth noting that in the capture of two electrons from
external atomic shells, the energy release is close to the energy
of the excited level of $^{190}$Os with $E_{exc}=1382.4$ keV
\cite{Sin03}. Such a coincidence could give a resonant
enhancement of the neutrinoless double electron capture in result
of energy degeneracy. The possibility of the resonant neutrinoless
double electron capture was discussed in Refs.
\cite{Win55,Vol82,Ber83}, where an enhancement of the rate by some
 orders of magnitude was predicted for the case of coincidence
between the released energy and the energy of an excited state.
According to \cite{Suj04}, high $Z$ atoms are strongly favored to
search for resonant $2\varepsilon$ decay. It should be noted that
$^{190}$Pt has the greatest $Z$ value among the nuclei for which a
resonant double electron capture is possible.

Unfortunately the natural abundance of $^{190}$Pt is very low:
$\delta=(0.014\pm0.001)\%$ \cite{Boh05}. Perhaps this fact, together
with the rather high cost of platinum, explains the absence of
experimental results (after the only one old study \cite{Fre52})
to search for $2\beta$ decay of this nuclide. A modest limit on
$\varepsilon\beta^+$ decay of $^{190}$Pt on the level of
$3.1\times10^{11}$ yr was calculated in \cite{DBD-tab}
on the basis of the early experiment \cite{Fre52} with corrections
on the decay energy, experimental efficiency and the natural isotopic abundance.

Another platinum isotope $^{198}$Pt is potentially $2\beta^{-}$
active, with the energy of decay $Q_{2\beta} =(1047\pm3)$ keV
\cite{Aud03}. The isotopic abundance of $^{198}$Pt is
$(7.163\pm0.055)\%$ \cite{Boh05}. The decay scheme of $^{198}$Pt
is presented in Fig. 2. As one can see, the $2\beta^-$ decay of
$^{198}$Pt to the first excited $2^+$ level of $^{198}$Hg can be
searched for by using an external $\gamma$ detector.

\begin{figure}[htb]
\begin{center}
\resizebox{0.4\textwidth}{!}{\includegraphics{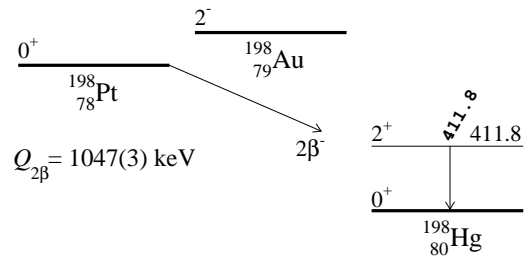}}
\end{center}
\caption{Decay scheme of $^{198}$Pt \cite{Chu02}. The energy of the 
excited level and of the $\gamma$ quantum are in keV. $Q_{2\beta}$ is the
double beta decay energy.}
\end{figure}

The aim of this study was the search for double $\beta$ processes in
$^{190}$Pt and $^{198}$Pt with the help of ultra-low background HP
Ge $\gamma$ spectrometry.

\section{Experiment}

Two platinum cups and one lid designed for chemistry purposes with
the total mass of 42.53 g were used in the experiment. Taking into
account the isotopic composition of the platinum, the sample contains
$1.84\times10^{19}$ nuclei of $^{190}$Pt and $9.40\times10^{21}$
nuclei of $^{198}$Pt. The search for double $\beta$ decay of
platinum was realized at the Laboratori Nazionali del Gran Sasso
of the INFN (average overburden of $\approx3600$ meters of water
equivalent) \cite{Arp02,Lau04} with a p-type HP Ge detector
(GeCris, 468 cm$^3$ of volume). The energy resolution of the
spectrometer is FWHM = 2.0 keV for the 1333 keV $\gamma$ line of
$^{60}$Co. The data with the sample were accumulated over 1815.4 h,
while the background spectrum was taken over 1045.6 h. The spectra
normalized on the time of measurements are presented in Fig.~3.

\begin{figure}[htb]
\begin{center}
\resizebox{0.7\textwidth}{!}{\includegraphics{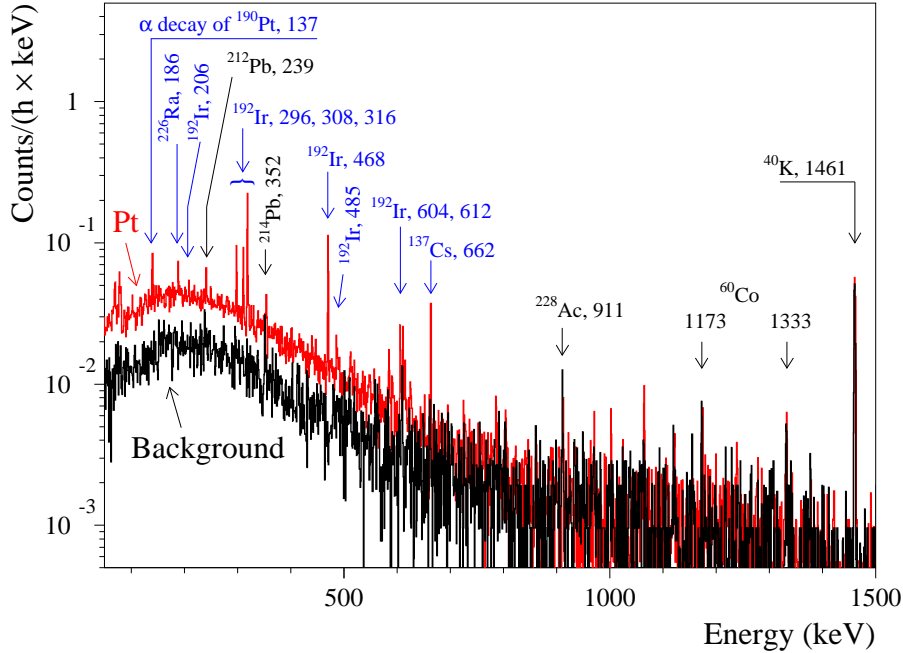}}
\end{center}
\caption{(Color online) Energy spectra measured with the 42.5 g platinum
sample over 1815 h (Pt) and without sample over 1046 h
(Background) by ultra-low background HP Ge $\gamma$ spectrometer.
The energy of the $\gamma$ lines are in keV.}
\end{figure}

\section{Results and discussion}

\subsection{Radioactive contamination of platinum}

There is a certain difference between the Pt and the background
spectra, mainly due to the contamination of the pla\-ti\-num by $^{192}$Ir
($T_{1/2}=73.831$ d \cite{ToI98}). The activity of $^{192}$Ir is
equal to ($40\pm5$) mBq/kg. The radioactive iridium could appear
in platinum due to the cosmogenic activation of Pt by cosmic rays at
the Earth surface. In addition, iridium usually accompanies
platinum in ores. Therefore $^{192}$Ir can be created in result of
neutron capture by $^{191}$Ir which is one of two naturally
occurring iridium isotopes ($\delta=37.3\%$, the cross section for
thermal neutrons is $\sigma=954$ b \cite{ToI98}). However, the
half-life of the ground state of $^{192}$Ir is $T_{1/2}=73.8$ d
\cite{ToI98}, and the exponential decrease in time of the
$^{192}$Ir activity should be observed during our 75.6 d
measurements. In fact, the behaviour of the counting rate in the
f.e. 468.1 keV $\gamma$ peak of $^{192}$Ir cannot be explained by
this assumption (see Fig. 4). A fit of the data by the exponential
function corresponding to the decay of $^{192}$Ir gives too large
value of $\chi^2$/n.d.f.$=22.6/9=2.5$. At the same time the data
is very well described ($\chi^2$/n.d.f.$=7.8/9=0.87$) by
exponential decay with the half-life 241 y (decay of $^{192m}$Ir,
see Fig. 4). Thus, the $^{192}$Ir activity should be ascribed not
to the ground state, but to the isomeric $^{192m}$Ir level with
$E_{exc}=168.1$ keV and $T_{1/2}=241$ yr \cite{Bag98}. This
isomeric state decays to the ground state of $^{192}$Ir emitting
155.1 keV and 13.0 keV $\gamma$'s which are however strongly
converted to electrons (coefficients of conversion are equal
to $\alpha_{13}=57000$ and $\alpha_{155}=1026$ \cite{Bag98}). This
explains the absence of the 155.1 keV peak in our data.

\begin{figure}[htb]
\begin{center}
\resizebox{0.5\textwidth}{!}{\includegraphics{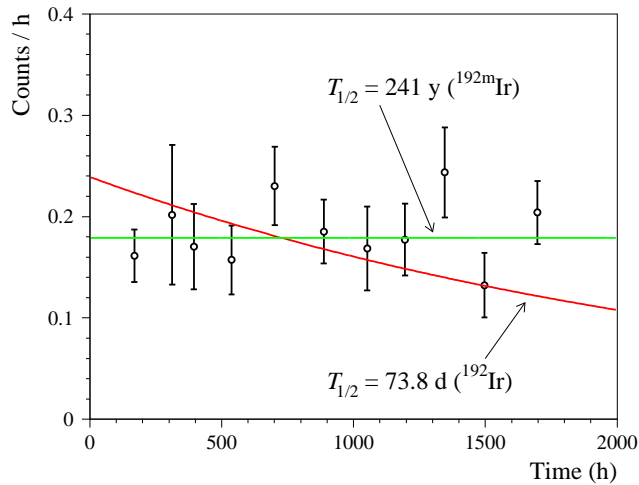}}
\end{center}
\caption{(Color online) Behaviour in time of the counting rate in
the 468.1 keV $\gamma$ peak from the decays of $^{192}$Ir in the
platinum sample measured by ultra-low background HP Ge $\gamma$
spectrometer. The dependence can be explained assuming the appearance
of $^{192}$Ir in the platinum sample after the decay of $^{192m}$Ir.
See text.}
\end{figure}

A peak at the energy 137 keV is caused by the $\alpha$ decay of
$^{190}$Pt to the lowest excited 137.2 keV level of $^{186}$Os
(the first evidence is reported in \cite{Bel10}). We have also
observed some excess of events in the 662 keV peak of $^{137}$Cs.
The response functions of the detector to decays of $^{40}$K,
$^{60}$Co, $^{137}$Cs, $^{192}$Ir, U/Th daughters in the platinum
sample were simulated by EGS4 code \cite{EGS4} with initial
kinematics given by the DECAY0 event generator \cite{DECAY0}. The
activity of $^{137}$Cs and $^{192}$Ir, as well as upper limits on
contamination by $^{40}$K, $^{60}$Co and  U/Th daughters are
presented in Table 1.

The investigation of radioactive contaminants in 
platinum samples is important also e.g. when selecting 
the materials to build platinum crucibles to be used for growing
radiopure inorganic crystal scintillators.

\begin{table}[htb]
\caption{Radioactive contamination of the platinum sample measured
with HP Ge detector. Gamma emitters and energies of $\gamma$ lines
used to determine the activity of the isotopes are specified in the 3rd
column. The upper limits are given at 90\% C.L., and the uncertainties
of the measured activities are estimated at 68\% C.L.}
\begin{center}
\begin{tabular}{llll}
\hline
 Chain      &  Nuclide      & $\gamma$ emitters         & Activity \\
 ~          & (Sub-chain)   & ($E_{\gamma}$, keV)       & (mBq/kg) \\
 \hline
 ~          & ~             & ~                         & ~  \\
 ~          & $^{40}$K      & (1460.8)                  & $\leq25$ \\
 ~          & $^{60}$Co     & (1173.2)                  & $\leq1.5$ \\
 ~          & $^{137}$Cs    & (661.6)                   & $7\pm1$  \\
 ~          & $^{192}$Ir    & (296.0, 316.5, 468.1)     & $40\pm5$ \\
 ~          & ~             & ~                         & ~  \\
 $^{232}$Th & $^{228}$Ra    & $^{228}$Ac (911.2)        & $\leq7$ \\
 ~          & $^{228}$Th    & $^{212}$Pb (238.6),       & ~ \\
 ~          & ~             & $^{208}$Tl (583.2, 2614.6)& $\leq7$  \\
 ~          & ~             & ~                         & ~  \\
 $^{235}$U  & $^{235}$U     & $^{235}$U (185.7)         & $\leq16$ \\
 ~          & $^{231}$Pa    & $^{231}$Pa (283.7, 300.1) & $\leq66$ \\
 ~          & $^{227}$Ac    & $^{227}$Th (236.0)        & $\leq13$ \\
 ~          & ~             & ~                         & ~ \\
 $^{238}$U  & $^{238}$U     & $^{234m}$Pa (766.4)       & $\leq68$ \\
 ~          & $^{226}$Ra    & $^{214}$Pb (352.0),       & ~ \\
 ~          & ~             & $^{214}$Bi (609.3, 1764.5)& $\leq3$ \\
  ~         & $^{210}$Pb    & $^{210}$Pb (46.5)         & $\leq34000$ \\
 ~          & ~             & ~                         & ~  \\
 \hline
\end{tabular}
\end{center}
\end{table}

\subsection{Search for double $\beta$ processes in $^{190}$Pt and $^{198}$Pt}

We do not observe any peaks in the spectrum accumulated with the
platinum sample which could indicate double $\beta$ activity of
$^{190}$Pt or $^{198}$Pt. Therefore only lower half-life limits
($\lim T_{1/2}$) can be set according to the formula: $\lim
T_{1/2} = N \cdot \eta \cdot t \cdot \ln 2 / \lim S$, where $N$ is
the number of potentially $2\beta$ unstable nuclei, $\eta$ is the
detection efficiency, $t$ is the measuring time, and $\lim S$ is
the number of events of the effect searched for which can be
excluded at given confidence level (C.L., all the limits obtained
in the present study are given at 90\% C.L.). The efficiencies of the
detector to the double $\beta$ processes in the platinum isotopes
were also calculated with the EGS4 code \cite{EGS4} and
DECAY0 event generator \cite{DECAY0}.

\begin{figure}[!ht]
\begin{center}
\resizebox{0.55\textwidth}{!}{\includegraphics{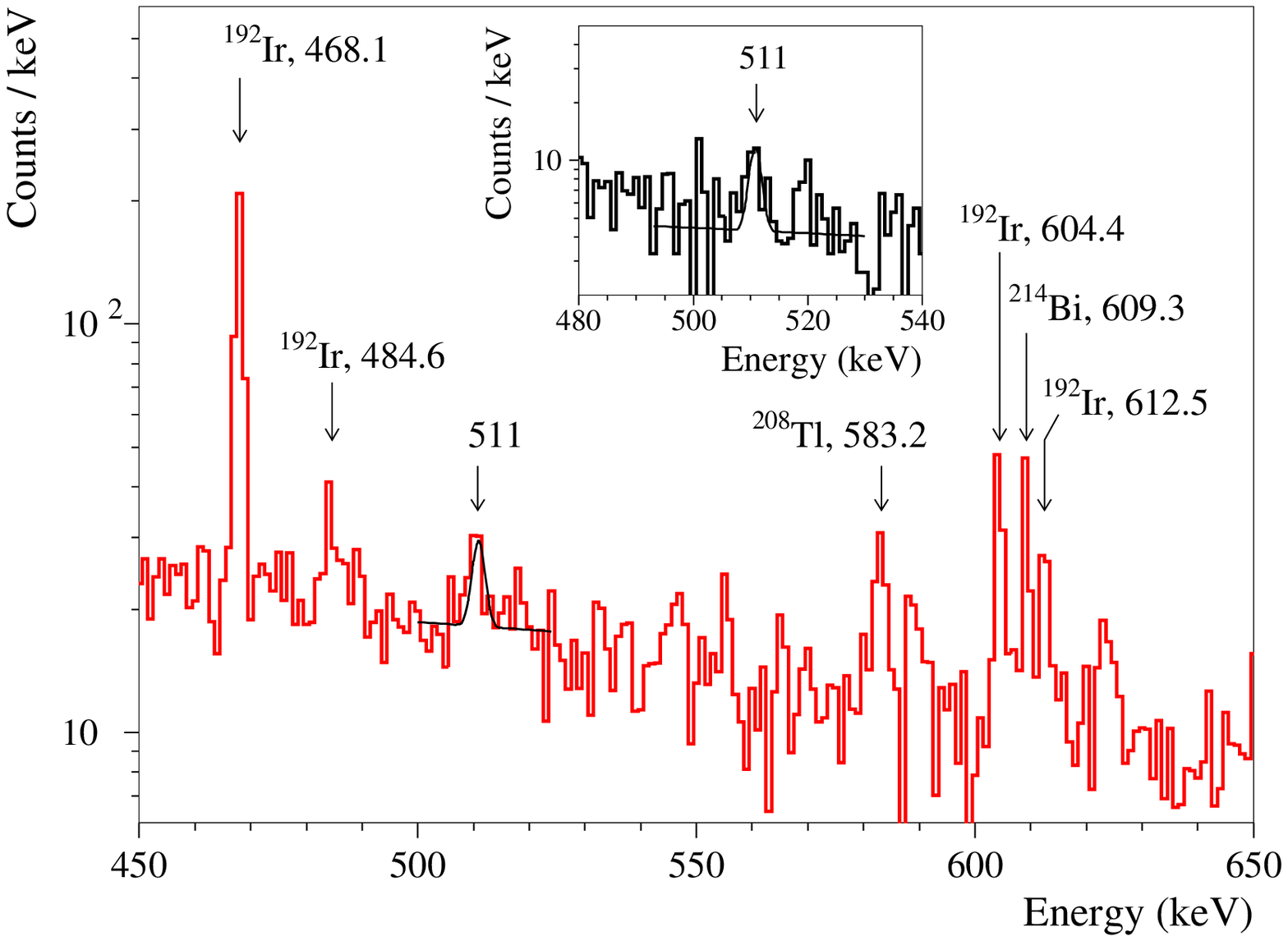}}
\end{center}
\caption{(Color online) Fragment of the energy spectra accumulated
with the platinum sample over 1815 h by ultra-low background HP Ge
$\gamma$ spectrometer. (Inset) The energy spectrum accumulated
without sample over 1046 h. The  fits of the 511 keV annihilation $\gamma$
peaks are shown by solid lines. The energy of the $\gamma$ lines are in
keV.}
\end{figure}

\subsubsection{Electron capture with positron emission in $^{190}$Pt}

One positron can be emitted in the $\varepsilon\beta^+$ decay of
$^{190}$Pt with energy up to ($361\pm6$) keV. The annihilation of
the positron will give two 511 keV $\gamma$'s leading to extra rate in
the annihilation peak.

Part of the spectrum in the energy interval $450-650$ keV is shown
in Fig. 5. There are annihilation peaks in both the spectra
accumulated with the platinum sample ($36\pm13$) counts and in the
background spectrum ($12\pm5$) counts. The decays of $^{137}$Cs and
$^{192}$Ir, present in the used platinum sample, provide no
contribution to the peak. The difference in the areas of the
annihilation peak ($15\pm16$) counts, which can be attributed to
electron capture with positron emission in $^{190}$Pt, gives no
indication on the effect. In accordance with the Feldman-Cousins
procedure \cite{Fel98} we should take $\lim S=41$ counts which can
be excluded at 90\% C.L. Taking into account the detection
efficiency ($\eta=14.2\%$), we have estimated a limit on the
half-life of $^{190}$Pt relatively to two neutrino
$\varepsilon\beta^+$ decay as:

\begin{center}
$T_{1/2}^{2\nu\varepsilon\beta^{+}}($g.s.$~\rightarrow~$g.s.$)\geq
9.2\times10^{15}$ yr.
\end{center}

The detection efficiency in a case of neutrinoless
$\varepsilon\beta^{+}$ decay is slightly lower (13.9\%), which
leads to the limit:

\begin{center}
$T_{1/2}^{0\nu\varepsilon\beta^{+}}($g.s.$~\rightarrow~$g.s.$)\geq
9.0\times10^{15}$ yr.
\end{center}

In the $2\nu\varepsilon\beta^+$ decay the first excited level of
$^{190}$Os could also be populated with emission of $\gamma$
quanta with the energy 186.7 keV. There is a peak at the energy of
($185.6\pm0.3$) keV with the area ($86\pm21$) counts in the spectrum
accumulated with the platinum sample (see Fig.~6, a). Most likely
this peak can be explained by $\alpha$ decay of $^{235}$U to the
excited levels of $^{231}$Th. Some part of the peak can
be due to $\alpha$ decay of $^{226}$Ra to the excited level 186.2
keV of $^{222}$Rn. At the same time we cannot surely estimate a
contribution to this peak neither from the decay of $^{235}$U nor
from the decay of $^{226}$Ra. For $^{235}$U we set only the limit on activity of
$^{235}$U in the platinum ($\leq16$ mBq/kg). Even much stronger
limit ($\leq3$ mBq/kg) was obtained on the activity of $^{226}$Ra
by analysis of $^{214}$Pb and $^{214}$Bi peaks. Thus, we ascribe
very conservately 120 counts to the $\varepsilon\beta^+$ decay of
$^{190}$Pt to the first excited level of $^{190}$Os, which leads
to the following limit on the process:
$T_{1/2}^{2\nu\varepsilon\beta^{+}}($g.s.$~\rightarrow
186.7~$keV$)\geq~9.0\times10^{14}$ yr. However, a better limit can
be set by using the estimation of the annihilation peak area (the
detection efficiency for the process is $\eta=13.0\%$):

\begin{center}
$T_{1/2}^{2\nu\varepsilon\beta^{+}}($g.s.$~\rightarrow
186.7~$keV$)\geq~8.4\times10^{15}$ yr.
\end{center}

\begin{figure}[!ht]
\begin{center}
\resizebox{0.5\textwidth}{!}{\includegraphics{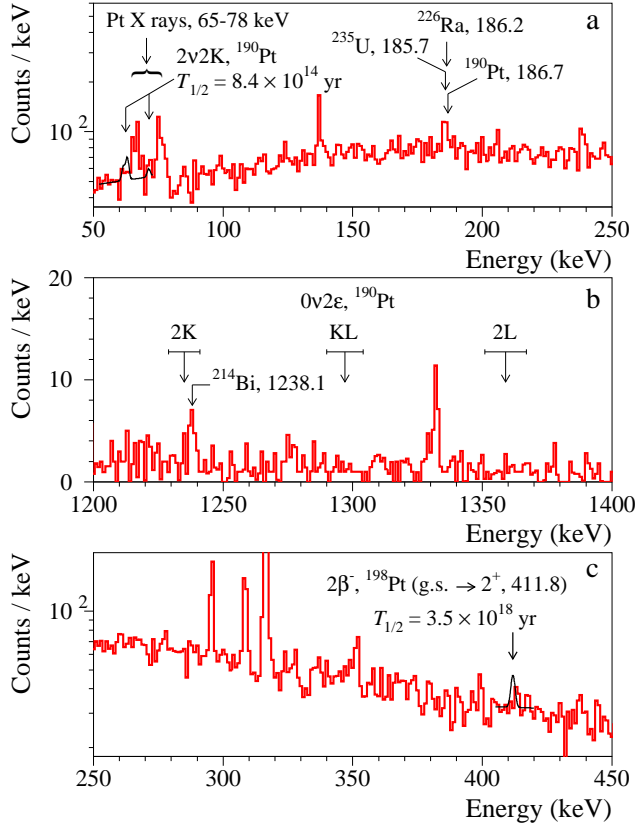}}
\end{center}
\caption{(Color online) (a) Part of the energy spectrum
accumulated with the platinum sample over 1815 h in the energy
interval ($50-250$) keV. The distribution expected for $2\nu2K$
decay of $^{190}$Pt with the half-life $8.4\times10^{14}$ yr
excluded at 90\% C.L. is shown by solid line. (b) Part of the Pt
energy spectrum where peaks from the $0\nu2\varepsilon$ processes in
$^{190}$Pt ($2K$, $KL$ and $2L$) to the ground state of $^{190}$Os
are expected. (c) Part of the spectrum in the energy interval
($250-450$) keV where a $\gamma$ peak with the energy of 411.8 keV
is expected for $2\beta^-$ decay of $^{198}$Pt to the excited
level $2^+_1$ of $^{198}$Hg. Area of the peak shown by solid line
corresponds to the half-life $3.5\times10^{18}$ yr excluded at
90\% C.L.}
\end{figure}

The same limit was set for the neutrinoless $\varepsilon\beta^+$
decay of $^{190}$Pt to the excited 186.7 keV level of $^{190}$Os.

\subsubsection{Double electron capture in $^{190}$Pt}

The $2\nu2K$ capture in $^{190}$Pt to the ground state of
$^{190}$Os will give cascade of X rays and Auger electrons with
the individual energies up to 73.8 keV \cite{ToI98}. The most
intensive X ray lines should have the energies 61.5 (27.4\%), 63.0
(46.7\%), 71.1 (5.5\%), 71.4 (10.6\%) and 73.4 keV (3.7\%).
Contributions to the spectrum in this energy region (see Fig.~6,
a) are expected from the decay of $^{192}$Ir, the excitation of Pt by
the decays of $^{137}$Cs, and also from the decays of U/Th daughters
contained in the materials of the detector. To set a limit on the
process searched for, the energy spectrum was fitted by the model
consisting of the sum of five Gaussians (to describe the expected
61.5, 63.0, 71.1, 71.4 and 73.4 keV peaks in $^{190}$Os), two
Gaussians (65.1 and 66.8 keV X ray lines due to excitation of Pt),
and a polynomial function of the first degree (background). A fit
in the energy interval ($52-73$) keV gives the area of the $2\nu2K$
effect as ($26\pm21$) counts. According to the Feldman-Cousins
procedure we should take $\lim S=60$ counts at 90\% C.L. Therefore
taking into account the efficiency to detect the expected effect
(1.9\%) we set the following limit on the $2\nu2K$ capture in
$^{190}$Pt:

\begin{center}
$T_{1/2}^{2\nu2K}($g.s.$~\rightarrow~$g.s.$)\geq~8.4\times10^{14}$
yr.
\end{center}

The distribution of the excluded effect of the $2\nu2K$ decay in
$^{190}$Pt is shown at Fig. 6, a.

The $Q_{2\beta}$ energy of $^{190}$Pt allows also the population of
several excited levels of $^{190}$Os. In a subsequent
de-excitation process, one or few cascade $\gamma$ quanta
(together with conversion electrons and $e^+e^-$ pairs) will be
emitted. To estimate limits on two neutrino double electron
capture in  $^{190}$Pt, the energy spectrum accumulated with the
platinum sample was analyzed in different energy intervals. The
results of the estimations are presented in Table 2, where the
energies of the $\gamma$ peaks, used to derive the limits, are also
specified.

In the neutrinoless double electron capture to the gro\-und state of
the daughter nuclei, in addition to the X rays, some other particle(s)
should be emitted to take away the rest of the energy. Usually one
bremsstrahlung $\gamma$ quantum is assumed. The energy of the
$\gamma$ quantum is expected to be equal to
$E_\gamma=Q_{2\beta}-E_{b1}-E_{b2}$, where $E_{b1}$ and $E_{b2}$
are the binding energies of the first and of the second captured
electrons on the atomic shell. The binding energies on the $K$,
$L_1, L_2$ and $L_3$ shells in Os are equal to $E_K=73.9$
keV, $E_{L_1}=13.0$ keV, $E_{L_2}=12.4$ keV and $E_{L_3}=10.9$ keV
\cite{ToI98}, respectively. Therefore, the expected energies of
the $\gamma$ quanta for the $0\nu2\varepsilon$ capture in
$^{190}$Pt to the ground state of $^{190}$Os are in the intervals:
i)$E_\gamma=(1229-1241)$ keV for the $0\nu 2K$; ii)
$E_\gamma=(1290-1304)$ keV for the $0\nu KL$; iii)
$E_\gamma=(1351-1367)$ keV for the $0\nu 2L$ process.

\begin{table}[htbp]
\caption{Half-life limits on 2$\beta$ processes in $^{190}$Pt and
$^{198}$Pt isotopes. The energies of the $\gamma$ lines
($E_\gamma$), which were used to set the $T_{1/2}$ limits, are
listed in column 4 with the corresponding efficiencies ($\eta$) in
column 5. $T_{1/2}$ limits are derived in the present work at 90\%
C.L., while the limit from \cite{DBD-tab,Fre52} is given at 68\%
C.L.}
\begin{center}
\resizebox{1.00\textwidth}{!}{
\begin{tabular}{lllllll}

 \hline
 Process                   & Decay       & Level of      & $E_\gamma$  & $\eta$ & \multicolumn{2}{c}{$T_{1/2}$ (yr)}  \\
 \cline{6-7}
 of decay                  & mode        & daughter      & (keV)       &        & Present work        & \cite{DBD-tab,Fre52} \\
 ~                         & ~           & nucleus       &             &        & ~                   & ~ \\
 ~                         & ~           & (keV)         &             &        &                     & ~ \\
 \hline
 ~                         & ~           & ~             & ~           & ~      & ~                   & ~  \\
 $^{190}$Pt$\to$$^{190}$Os & ~           & ~             & ~           & ~      & ~                   & ~  \\
 $\varepsilon\beta^+$      & $2\nu$      & g.s.          & 511         & 14.2\% & $>9.2\times10^{15}$ & -- \\
 ~                         & $0\nu$      & g.s.          & 511         & 13.9\% & $>9.0\times10^{15}$ & $>3.1\times10^{11}$  \\
 ~                         & $2\nu+0\nu$ & $2^+$ 186.7   & 511         & 13.0\% & $>8.4\times10^{15}$ & -- \\
 ~                         & ~           & ~             & ~           & ~      & ~                   & ~  \\
 $2K$                      & $2\nu$      & g.s.          & $61.5-73.4$ & 1.9\%  & $>8.4\times10^{14}$ & -- \\
 ~                         & $2\nu$      & $2^+$ 186.7   & $61.5-73.4$ & 2.0\%  & $>8.8\times10^{14}$ & -- \\
 ~                         & $2\nu$      & $2^+$ 558.0   & 558.0       & 4.0\%  & $>4.0\times10^{15}$ & -- \\
 ~                         & $2\nu$      & $0^+$ 911.8   & 725.1       & 4.4\%  & $>4.5\times10^{15}$ & -- \\
 ~                         & $2\nu$      & $2^+$ 1114.7  & 1114.7      & 1.6\%  & $>1.0\times10^{16}$ & -- \\
 $2K$                      & $0\nu$      & g.s.          & $1229-1241$ & 4.7\%  & $>5.7\times10^{15}$ & -- \\
 $KL$                      & $0\nu$      & g.s.          & $1290-1304$ & 4.6\%  & $>1.7\times10^{16}$ & -- \\
 $2L$                      & $0\nu$      & g.s.          & $1351-1367$ & 4.6\%  & $>3.1\times10^{16}$ & -- \\
 $2\varepsilon$            & $0\nu$      & $2^+$ 186.7   & 186.7       & 3.1\%  & $>6.9\times10^{14}$ & -- \\
 ~                         & $0\nu$      & $2^+$ 558.0   & 558.0       & 3.1\%  & $>4.5\times10^{15}$ & -- \\
 ~                         & $0\nu$      & $0^+$ 911.8   & 725.1       & 3.6\%  & $>3.6\times10^{15}$ & -- \\
 ~                         & $0\nu$      & $2^+$ 1114.7  & 1114.7      & 1.5\%  & $>9.8\times10^{15}$ & -- \\
 ~                         & ~           & ~             & ~           & ~      & ~                   & ~  \\
 Resonant $MM$, $MN$, $NN$ & $2\nu+0\nu$ & $(0,1,2^+)$ 1382.4 & 1195.7 & 4.5\%  & $>2.9\times10^{16}$ & -- \\
 ~                         & ~           & ~             & ~           & ~      & ~                   & ~  \\
 $^{198}$Pt$\to$$^{198}$Hg & ~           & ~             & ~           & ~      & ~                   & ~  \\
 $2\beta^-$                & $2\nu+0\nu$ & $2^+$  411.8  & 411.8       & 7.5\%  & $>3.5\times10^{18}$ & -- \\

 \hline

 \end{tabular}
 }
 \end{center}
 \end{table}

There is only one peak at the energy of 1238 keV with the area
($15\pm4$) counts in the spectrum accumulated with the platinum
sample in the energy interval of interest ($1229-1241$) keV (see
Fig.~6, b). More likely this is the $\gamma$ peak 1238.1 keV from the
decays of $^{214}$Bi (daughter of $^{226}$Ra from $^{238}$U
family). It is rather difficult to estimate a contribution of the
$^{214}$Bi $\gamma$ quanta to the peak area because we do not know
exactly location of the $^{226}$Ra contamination in the materials
of the set-up. Conservatively we assume that all the peak area is due
to the neutrinoless $2K$ decay of $^{190}$Pt. Taking into account
the calculated efficiency to detect $\gamma$ quanta with the
energy ($1229-1241$) keV ($\approx4.7\%$), it gives the following
limit on the process searched for:

 \begin{center}
 $T_{1/2}^{0\nu2K}($g.s.$~\rightarrow~$g.s.$)\geq~5.7\times10^{15}$ yr.
 \end{center}

There are no clear peaks in the energy intervals ($1290-1304$) keV
and ($1351-1367$) keV expected for the $0\nu KL$ and the $0\nu 2L$ processes
in $^{190}$Pt. Taking into account the calculated detection
efficiencies for $\gamma$ quanta with the energies in these
intervals (4.6\%) and limits on the numbers of events of the
effects ($\lim S=7$ and $\lim S=3.9$ counts, respectively) we have
obtained the following limits on the $0\nu$ double electron
captures in $^{190}$Pt to the ground state of $^{190}$Os:

 \begin{center}
 $T_{1/2}^{0\nu KL}($g.s.$~\rightarrow~$g.s.$)\geq~1.7\times10^{16}$ yr,
 \end{center}

 \begin{center}
 $T_{1/2}^{0\nu2L}($g.s.$~\rightarrow~$g.s.$)\geq~3.1\times10^{16}$ yr.
 \end{center}

Limits on the $0\nu2\varepsilon$ decay of $^{190}$Pt to the
excited levels of $^{190}$Os were obtained in the similar way as
for the $2\nu$ mode by analysis of the experimental data in the
energy intervals where intensive $\gamma$ peaks from the processes
are expected.

The limits on double electron capture in $^{190}$Pt to the ground
and excited levels of $^{190}$Os are presented in Table~2.

\subsubsection{Resonant $2\varepsilon$ capture in $^{190}$Pt}

The atomic mass difference between $^{190}$Pt and $^{190}$Os atoms
is very close to the energy of the excited level 1382.4 keV in
$^{190}$Os. A resonance transition could be fulfilled if two
external electrons will be captured. There is no peak in the
energy spectrum accumulated with the platinum sample with the
energy 1195.7 keV expected from de-excitation of the 1382.4 keV
level of $^{190}$Os (see Fig.~1). Taking into account the detection efficiency
for 1195.7 keV $\gamma$ quanta (4.5\%) we set the following limit
on the decay:

 \begin{center}
 $T_{1/2}^{2\varepsilon(0\nu+2\nu)}($g.s.$~\rightarrow 1382.4$~keV$)\geq 2.9\times10^{16}$ yr.
 \end{center}

\subsubsection{Double $\beta^-$ decay of $^{198}$Pt}

To set a limit on the double $\beta^-$ transition of $^{198}$Pt to
the $2^+_1$ excited level of $^{198}$Hg with the energy of 411.8
keV, the energy spectrum (see Fig. 6, c) was fitted in the energy
interval ($405-420$) keV by a straight line (which represents the
background model) and the expected peak at 411.8 keV. The fit
gives the area of the peak ($13\pm10$) counts, which allows to
exclude 29 counts at 90\% C.L. Taking into account the detection
efficiency ($\eta=7.5\%$) we have obtained the following limit on
the process searched for:

 \begin{center}
 $T_{1/2}^{2\beta^-(2\nu+0\nu)}($g.s.$~\rightarrow 411.8$~keV$)\geq 3.5\times10^{18}$ yr.
 \end{center}

\section{Conclusions}

The measurements performed over 1815 h with a 42.5 g sample of
platinum with the help of an ultra-low background HP Ge $\gamma$
spectrometer (468 cm$^3$) were used to set limits on double
$\beta$ processes in $^{190}$Pt in the range of $T_{1/2}\sim
10^{14-16}$ yr, which is $3-4$ orders of magnitude higher than the
limit on $0\nu\varepsilon\beta^+$ (g.s. $\to$ g.s.) decay obtained
in \cite{DBD-tab} by a re-analysis of the early experiment
\cite{Fre52}.

The search for the possible resonant $0\nu2\varepsilon$ capture to the
1382.4 keV level was realized for the first time. For future applications, 
taking into account the strong dependence of the resonant process on the
difference in atomic masses of $^{190}$Pt and $^{190}$Os,
precise measurements of the atomic masses are required. Moreover, it would
also be useful to precisely study the characteristics of the 1382.4
keV level of $^{190}$Os (spin, parity, decay scheme).

Search for $2\beta$ transition of $^{198}$Pt to the 411.8 keV
excited level of $^{198}$Hg was carried out at the first time
giving the limit $T_{1/2}>3.5\times 10^{18}$ yr.

The measurements allowed us to estimate the radioactive contamination
of the used platinum sample. In particular, we have detected 7~mBq/kg of $^{137}$Cs and
40~mBq/kg of $^{192}$Ir in this platinum sample. The contamination
of $^{60}$Co, $^{226}$Ra, $^{228}$Ra and $^{228}$Th does not
exceed the level of a few mBq/kg, while the activities of
$^{40}$K, $^{235}$U, $^{238}$U are less than a few tens mBq/kg (we
assume a broken equilibrium in U/Th chains). 

To improve the sensitivity of the experiment we are going to
increase the mass of the platinum by one order of magnitude. Further
improvements can be achieved by increasing the detection efficiency
and the exposition, and obviously by using enriched $^{190}$Pt
isotope. Specially developed multi-crystal HP Ge detectors could
also be applied to reach a sensitivity to double $\beta$ processes in
$^{190}$Pt on the level of $10^{22-24}$ yr. This would be
particularly interesting also because of the possibility of a
resonant double electron capture in $^{190}$Pt having such isotope 
the largest $Z$ value (the nuclear parameter which greatly favors the
process) among the nuclei where a resonant double electron
capture could occur.

\section{Acknowledgements}

The group from the Institute for Nuclear Research (Kyiv, Ukraine)
was supported in part through the Project
``Kos\-mo\-mikro\-fizyka-2'' (Astroparticle Physics) of the
National Academy of Sciences of Ukraine.

\end{document}